\documentclass[11pt]{article}

\newcommand{\blind}{0}

\usepackage{natbib}            
\usepackage{amsmath, amsfonts, amssymb}   
\usepackage{comment}
\usepackage{amsthm}                     

\usepackage{color}             

\usepackage{pdflscape}         
\usepackage{graphicx}          

\usepackage{algorithm}   
\usepackage{algorithmic}

\newcommand{\bd}{\boldsymbol}
\newcommand{\ud}{\underline}
\newcommand{\bdud}[1]{  \ensuremath{ \ud{\bd{#1}} }  }
	
\newcommand{\defi}{\stackrel{\text{def}}{=}}	
\newcommand{\first}[2]{  \ensuremath{ \frac{\partial   #1}{\partial #2  } }  }

\newcommand{\ttplus}{\text{\sc{e}}}

\newcommand{\Mtot}{M_{\text{tot}}}

\newcommand{\qqq}{\textcolor{red}{ \underline{\textbf{???}} }}

\begin{document}

\def \spacingset#1{ \renewcommand{\baselinestretch}{#1}\small\normalsize } \spacingset{1}

\if0\blind
{
  \title{\bf Bayesian Calibration of Computer Models with Informative Failures}
  \author{Peter W. Marcy\thanks{
     This work was funded by the U.S. Department of Energy, Office of Fossil Energy's Carbon Capture Simulation Initiative through the National Energy Technology Laboratory.  The authors thank Canhai (Kevin) Lai and Zhijie (Jay) Xu for providing the MFIX simulator data as well as Derek Bingham for mentioning the COMPAS model.  
    The authors are also grateful to three anonymous reviewers for their helpful comments. }\hspace{.2cm} \\
    Statistical Sciences Group (CCS-6), Los Alamos National Laboratory \\
    and \\
    Curtis B. Storlie \\
    Mayo Clinic}
   \date{  \vspace*{-30pt}}
  \maketitle
} \fi

\if1\blind
{
  \bigskip
  \bigskip
  \bigskip
  \begin{center}   {\LARGE\bf Bayesian Calibration of Computer Models with Informative Failures}   \end{center}
  \medskip
} \fi

\bigskip
\begin{abstract}
There are many practical difficulties in the calibration of computer models to experimental data.  
%
%
One such complication is the fact that certain combinations of the calibration inputs can cause the code to output data lacking fundamental properties, or even to produce no output at all.
In many cases the researchers want or need to exclude the possibility of these ``failures" within their analyses.
We propose a Bayesian (meta-)model in which the posterior distribution for the calibration parameters naturally excludes regions of the input space corresponding to failed runs.  
That is, we define a statistical selection model to rigorously couple the disjoint problems of binary classification and computer model calibration.  
We demonstrate our methodology using data from a carbon capture experiment in which the numerics of the computational fluid dynamics are prone to instability.

\end{abstract}

\noindent%

\vfill

\newpage
\spacingset{1.45}     

\spacingset{1.25}     

\section{Introduction}

Computer codes (also called \emph{simulators}) are often used in the physical sciences to model complex systems and phenomena.  
Beginning with \cite{Sacks1989}, the statistical design/analysis of experiments involving simulators has become so ubiquitous that this subdiscipline needs little introduction; uninitiated readers may consult the plethora of standard references within \cite{Fang2006}, \cite{Levy2010}, and \cite{Santner2018}.

One typical inferential goal involving computer models is \emph{calibration}:
when actual experimental/field data of a system are available these measurements can be used to infer the most probable values of unobservable input quantities necessary to run the computer model.
Historically, calibrating computer models to data took the form of nonlinear regression \citep[e.g.]{Seber1989} in which a least-squares criterion was minimized, an early example being \cite{Box1956}.
Modern calibration analyses often feature fast approximations of the computer model and can go by the names ``tuning" \citep{Park1991, Cox2001} or ``history-matching" \citep{Craig1997}.
The latter takes a Bayes-linear approach and explicitly incorporates a discrepancy term to acknowledge the imperfection of the computer model.
Bayesian calibration featuring Gaussian process (GP) surrogates (\emph{emulators}) and model discrepancy can be found in \cite{Kennedy2001}, \cite{Higdon2008}, and a glut of subsequent modifications/applications by many authors.

One impediment to calibration, or even emulation of the computer model, is the possibility that certain combinations of the calibration inputs will result in output absent some necessary property $\mathcal P$: in other words, a failure.
In certain cases the scientists consider these failures data;
as such, the parameter combinations leading to failures are deemed \emph{implausible}, to use the language of history-matching.
When this is true, the failed runs should be used to exclude or down-weight those regions of the calibration input space. 
Without doing so it is possible that these regions will be admitted to the calibration posterior via an emulator which is forced to extrapolate where there is no data.
A smaller parameter space could be desirable in its own right, but especially when the calibrated model is to be used for further purposes such as uncertainty analysis \citep{Oakley2002}.

Here is some concrete motivation by way of examples.
First consider the Allometrically Constrained Growth and Carbon Allocation model of
\cite{Gemoets2013} and \cite{Fell2018}.
Calibrating this code to a particular tree using radius or height is bedeviled by the fact that the model will sometimes report the tree as being dead when the real tree is very much alive (Property $\mathcal P$).
As the code behaves this way for particular and a priori unknown calibration input configurations, it would be all too reasonable to estimate and utilize the $\mathcal P$-constraint on the parameter space.
(If the constraint were \emph{known}, this could be easily accommodated by the implausibility metric for a history-match.)
Another example 
is the COMPAS (Compact Object Mergers: Population Astrophysics and Statistics)
model of binary black hole formation \citep{Barrett2016}.
Calibrating this code to existing binary black holes using chirp mass only makes sense when ($\mathcal P$:) a black hole is actually formed-- this is not guaranteed for all inputs to the code.
%
%
The motivation behind our work is the calibration of data to a particular computational fluid dynamics (CFD) model.
The code is prone to numerical instability for certain parts of the input space, but its issues are not easily addressed as it is specialized software.
Property $\mathcal P$ in this case is the very presence of output.
Those regions of parameter space yielding no output pragmatically needed to be ruled out for further upscaling and uncertainty analyses because these were built around the same code.

The outline of the paper is as follows.
In Sections \ref{sec:GPcal} and \ref{sec:GPbin} we review the disparate methodologies of calibration and classification upon which our method will build.
In Section \ref{sec:GPboth} we combine them.
The new methodology will be demonstrated using the CFD code mentioned above with data from a carbon capture experiment in Section \ref{sec:Examples}.

\section{Computer Model Calibration Using Gaussian Processes}   \label{sec:GPcal}

In this section we review the details of univariate calibration using the framework of \cite{Kennedy2001}.  
The restriction to a single output is purely for expositional clarity-- because our method uses only the input space to couple the two sources of information, it is directly applicable to the case of multivariate output.

The inputs to the computer model are partitioned into two groups.  
The \emph{calibration} inputs $\bd t = (t_1,\ldots,t_{D_t})$ are quantities that can be changed within the code, but ideally would be fixed at one setting $\bd\theta = (\theta_1,\ldots,\theta_{D_t})$ that represents the true state of nature.  
These calibration parameters $\bd\theta$ represent physical quantities of interest that cannot directly be measured in the real system.  
The second group contains the \emph{variable} inputs $\bd x = (x_1,\ldots,x_{D_x})$ that correspond to observable or controllable experimental settings.
The computer code $\eta$ is executed $M$ times to produce simulated data $\eta_m = \eta(\bd x^*_m , \bd t^*_m), \ (m=1,\ldots,M)$.  
In addition there are $N$ measurements of the true physical process $\tau$:
\begin{align*}
      y_n   &=   \tau(\bd x_n) + \epsilon_n   \quad \quad   n = 1,\ldots,N \ ,
\end{align*}
where $\{ \epsilon_n \}_{n=1}^N$ are $iid$ $N(0,\sigma^2_\epsilon)$ observational errors.  
Note that this experimental data is the result of conditions $\bd x_n$ which may differ from those of the simulated data (hence the ``$*$" notation above).

The end goal is to use the simulated data to find the values of $\bd\theta$ most consistent with the experimental data.  
To make this inference the following relationship is assumed:
\begin{align*}
   \tau(\bd x)   &=   \eta(\bd x, \bd\theta) + \delta(\bd x) \ .
\end{align*}
That is, the real physical system can be decomposed as the simulated system at a ``true" setting $\bd\theta$, plus a systematic discrepancy which depends only on the variable inputs.  
\cite{Kennedy2001} use independent GPs to model the unknown functions $\eta(\bd x, \bd t)$ and $\delta(\bd x)$ so that uncertainty at unobserved input configurations can be accounted for \citep[e.g.,][]{OHagan1978, Sacks1989, Santner2018}.  
Specifically, $\eta$ is assumed to have mean $\mu_\eta$ and covariance kernel $k_\eta \left( (\bd x_1,\bd t_1), (\bd x_2,\bd t_2) \right)$; the discrepancy $\delta$ is assumed to have mean $\mu_\delta$ and covariance kernel $k_\delta \left( \bd x_1, \bd x_2 \right)$.  
In this work, we take the kernels of the independent processes to have the form:
\begin{align*}
   k_\eta \left( (\bd x_1,\bd t_1), (\bd x_2,\bd t_2) \right)   &=   \sigma^2_\eta   \prod_{d=1}^{D_x} R \left( |x_{1,d}-x_{2,d}| ; \lambda_{\eta,x,d} \right)   \cdot   \prod_{e=1}^{D_t} R \left( |t_{1,e}-t_{2,e}| ; \lambda_{\eta,t,e} \right)   \\
   k_\delta \left( \bd x_1, \bd x_2 \right)   &=   \sigma^2_\delta   \prod_{d=1}^{D_x} R \left( |x_{1,d}-x_{2,d}| ; \lambda_{\delta,d} \right)  \ ,   
\end{align*} 
where $R(\cdot \ ; \lambda)$ is a one-dimensional correlation function (such as a squared-exponential or Mat\'{e}rn with specified smoothness) and $(\bd\lambda_{\eta,x}, \bd\lambda_{\eta,t}, \bd\lambda_\delta) \defi \bd\lambda$ are $2D_x+D_t$ correlation length parameters.
Under these assumptions, together with the relationships
\begin{align}
   y_n   &=   \eta(\bd x_n, \bd\theta) + \delta(\bd x_n) + \epsilon_n   \quad \quad   n = 1,\ldots,N   \label{eq:DataCal} \\
   \eta_m   &=   \eta(\bd x^*_m , \bd t^*_m)   \hspace*{1.12in}   m=1,\ldots,M \ ,   \nonumber  
\end{align}
the joint likelihood for all of the observations is obtained readily.  
Let $\bd y$ denote all the experimental data and $\bd\eta$ denote the collected simulation runs.  
The distribution for all the observed data $\bd d \defi (\bd y^\top , \bd\eta^\top)^\top$ is then multivariate normal (MVN):
\small
\begin{align}
   N_{N+M} \Bigg( 
\text{mean }   &=   \left[\begin{array}{r} (\mu_\eta+\mu_\delta) \bd 1_N \\ \mu_\eta \bd 1_M \end{array}\right] ,   \nonumber  \\  
\text{cov }  &= 
\left[\begin{array}{l|c} 
   k_\eta([\bd X , \bdud\theta],[\bd X , \bdud\theta]) + k_\delta(\bd X,\bd X) + \sigma^2_\epsilon\bd I_N  &  k_\eta([\bd X , \bdud\theta],[\bd X^* , \bd T^*]) \\ 
   \hline k_\eta( [\bd X^* , \bd T^*],[\bd X , \bdud\theta]) & k_\eta( [\bd X^* , \bd T^*],[\bd X^* , \bd T^*]) \end{array}\right]
   \Bigg)   \ .  \label{eq:MVNcal}
\end{align}
\normalsize
Above, $k_\eta( [\bd X^* , \bd T^*],[\bd X , \bdud\theta])$ is the matrix whose $(i,j)$-entry is $k_\eta \left( (\bd x^*_i,\bd t^*_i), (\bd x_j,\bd\theta) \right)$;
other blocks within the covariance are defined similarly.  
If the non-calibration parameters are collected in the vector $\bd\alpha_{\eta , \delta} \defi (\mu_\eta, \mu_\delta, \ \sigma^2_\eta, \sigma^2_\delta, \sigma^2_\epsilon, \bd\lambda)$, the posterior distribution for all parameters is
\begin{align}
   [\bd\theta , \bd\alpha_{\eta , \delta} \ | \bd d]   &\propto   L(\bd\theta , \bd\alpha_{\eta , \delta} \ ; \bd d) \cdot [\bd\theta | \bd\alpha_{\eta , \delta}] \cdot [\bd\alpha_{\eta , \delta}]   \label{eq:PostCal} \ ,
\end{align}
and \emph{a priori} independence is almost always assumed: $[\bd\theta | \bd\alpha_{\eta , \delta}]  \equiv  [\bd\theta]$.
The notation $[X]$ is shorthand for $f_X(x)$, the density or mass function of the variable $X$, and $L(\cdot \ ; \bd d)$ represents the likelihood function which in this case is proportional to the MVN pdf defined by (\ref{eq:MVNcal}).  
We have partitioned the parameter space into emulator/discrepancy and calibration parameters to make later sections more transparent.

The posterior given in (\ref{eq:PostCal}) can be explored using Markov Chain Monte Carlo (MCMC) with Metropolis-Hastings updates for some of the parameters and Gibbs steps for the remaining.  
In updating the emulator/discrepancy parameters, the use of Gibbs versus Metropolis will depend on the prior specification.  
For all analyses in this paper we use the following prior distributions (which assume that the inputs within $\bd x$ and $\bd t$ have all been scaled to [0,1], and that the output $\bd d$ has been rescaled by its sample standard deviation):
\begin{align*}
   [\bd\theta | \bd\alpha_{\eta , \delta}] \cdot [\bd\alpha_{\eta , \delta}]   &\defi   [\bd\theta] \cdot [\bd\alpha_{\eta , \delta}]  \\
   &=   [\bd\theta]   \cdot   [\mu_\eta] \cdot [\mu_\delta]   \cdot    [\sigma^2_\eta] \cdot [\sigma^2_\delta] \cdot [\sigma^2_\epsilon]   \cdot   [\bd\lambda_{\eta,x} , \bd\lambda_{\eta,t} , \bd\lambda_\delta]  \\
   [\mu_\eta]\cdot[\mu_\delta]  &\propto  1  \\
   \sigma_\eta   &\sim   \text{Unif}(0,3)   \quad \quad   \sigma_\delta \sim \text{Unif}(0,2)   \quad \quad   \sigma_\epsilon \sim \text{Unif}(0,1)   \\
   \lambda   &\stackrel{iid}{\sim}   \text{Unif}(0.1, 5.0)   \quad \forall \ \lambda \in \{ \bd\lambda_{\eta,x} , \bd\lambda_{\eta,t} , \bd\lambda_\delta \}  \ . 
\end{align*}
The use of uniform priors on the square root of the variances was inspired by \cite{Gelman2006}, and we found that this produces better mixing than a typical inverse-gamma specification.  
All correlation functions in this work are parameterized such that a correlation length of $\lambda > 1$ implies a very smooth process;
the upper bound of 5.0 for the uniform priors on the $\lambda$'s works well to keep the covariance of the likelihood well-conditioned without sacrificing the emulator's predictive ability.  
Finally, the priors for the calibration parameters will depend on the particular application.  
Therefore, under this scheme of prior distributions, $\mu_\eta$ and $\mu_\delta$ can be updated by Gibbs, and the rest of the parameters by Metropolis steps.

\subsection{Illustrative Example (intro.)}

We now introduce a simple ``toy" example with which to illustrate our methodology in stages.  

\begin{figure}[h!t]   \begin{center}
\includegraphics[width=6.5in]{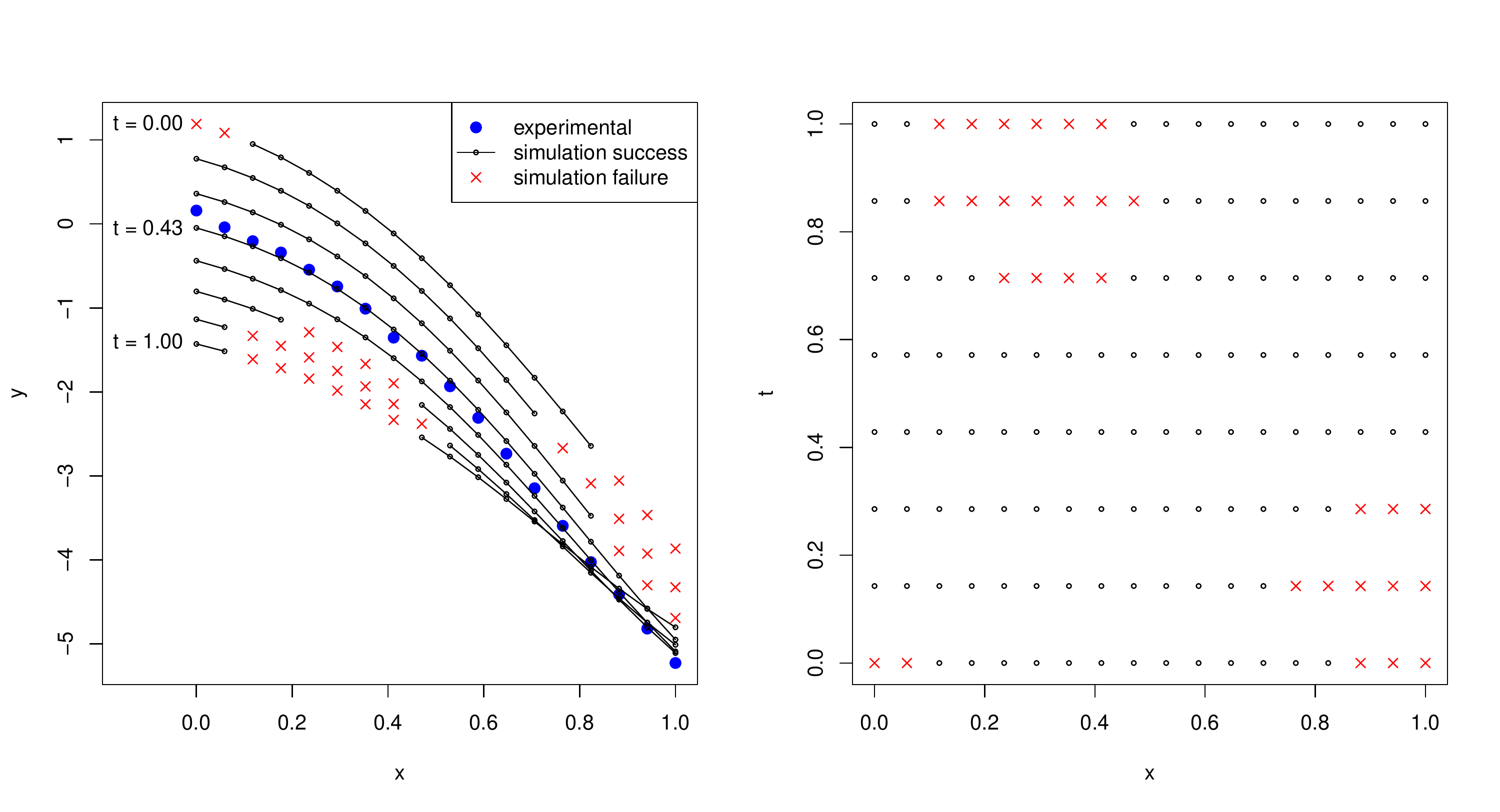}
\caption{ A simple calibration problem with failed ``simulator" runs.  The full data is given in the left panel, and the input space showing the failed runs is presented in the right panel. }
\label{fig:GPcal}
\end{center}   \end{figure}

\begin{figure}[h!t]   \begin{center}
\includegraphics[width=4.5in]{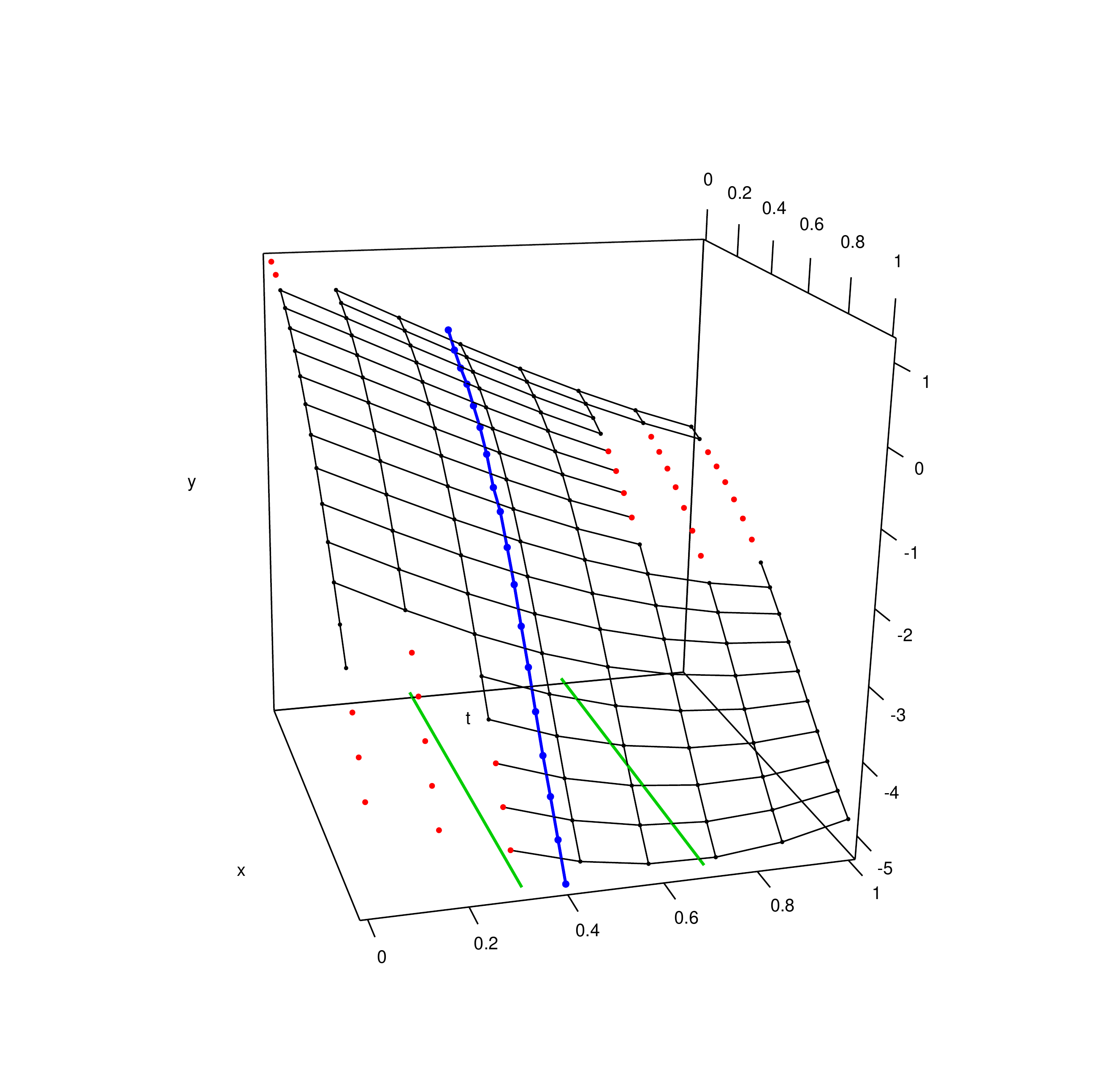}
\caption{ Response surface of completed ``simulator" runs for the illustrative example.  The experimental data (blue points) are plotted along the slice $t=\theta = 0.4$ and are connected by line segments. The green lines highlight the true region of $t$-space for which all runs are successful. }
\label{fig:GPcal3D}
\end{center}   \end{figure}

The data are obtained using (\ref{eq:DataCal}), where the ``simulator" $\eta(\cdot,\cdot)$ in this case is actually a realization of a zero-mean GP on the domain $(x,t) \in [0,1]^2$ having covariance kernel
$$ k_\eta( (x_1,t_1),(x_2,t_2) ) = \sigma^2_\eta  \exp \left( (x_1 - x_2)^2 /\lambda_{\eta , x}^2 \right) \cdot \exp \left( (t_1 - t_2)^2 /\lambda_{\eta , t}^2 \right) $$
with parameter values $\sigma^2_\eta=10$, $\lambda_{\eta , x}=1.0$, and $\lambda_{\eta , t}=2.0$.  
The discrepancy function is taken to be $\delta(x) = 0.1 (x-0.2)^2 - 0.5 (x-0.2)$ and the experimental error variance to be $\sigma^2_\epsilon = 0.002$.  
The hypothetical simulator runs are obtained on a $18 \times 8$ grid of the unit square, but we suppose that 30 of these points result in failures of some kind (so $M=114$).  
The experimental data of the physical system are obtained at the same $x$ values as the simulated data ($\bd X^* = \bd X$, N=18) using the true value of $t$ set to $\theta = 0.4$.

The successful simulations are shown in the left panel of Figure \ref{fig:GPcal} (black points/curves) together with the experimental data (blue points); 
the pattern of the failed runs are shown in the right panel.  
A different view of both datasets is presented in Figure \ref{fig:GPcal3D}; the response surface is presented with missing regions corresponding to failed simulation runs. 
As a visual aid, the experimental data are connected by line segments and plotted at $t=\theta$ (of course, the goal of calibration is to infer this location).
The green lines indicate the true region of the calibration input space for which all runs are successful.

We set a uniform prior on $\theta$ and use a fairly standard MCMC routine to explore the posterior distribution of all parameters $(\theta , \bd\alpha_{\eta , \delta})$.
A total of 120,000 iterations (20,000 of those for burn-in) were completed, and every third draw was recorded.  
Diagnostics for all parameters within the thinned chain were deemed adequate.
A plot of the estimated posterior density for $\theta$ is given by the black curve in Figure \ref{fig:PosteriorCompare01}.
Here it can be seen that the tails of this distribution give positive probability to regions of $\theta$ that give rise to simulator failures for certain values within the $x$-space.
If this calibrated model were to be used in a subsequent uncertainty analysis, the additional support would add superfluous variation to the quantity of interest.

\section{A Binary Random Field Model for Success/Failure Data}   \label{sec:GPbin}

In addition to the completed computer model runs $\bd\eta$, we assume that there is also a set of $M_0$ input configurations that result in model failures; let these be denoted $(\bd x^*_m , \bd t^*_m)$, where $m=(M+1),\ldots,(M+M_0)$.  
\emph{Latent variable} approaches can be used to model the binary outcome of success or failure at arbitrary inputs $(\bd x , \bd t)$ given the $\Mtot \defi M+M_0$ observed outcomes. 
 
In this section we describe Bayesian estimation and prediction under one such model: the binary or ``clipped" random field. 
Within this model, a response variable realization is 1 if and only if a continuous latent random field is positive.
This differs from a more standard logistic/probit approach in which a transformed random field (or regression function) determines the non-binary probability that a response value will be 1 \citep{Chib1998, Neal1999, Rasmussen2006} instead of the outcome itself.
%
The binary random field model will lead to sharper inference and predictions when the successes/failures are strongly correlated in space.
We will indicate when and how this model can be extended.

\subsection{Bayesian Implementation}  \label{sec:GPbinMod}

We define an indicator random variable $Z_m$ to be 1 if and only if the $m$th run is successful, and this is tied to a latent continuous random variable $\zeta_m$ through the relationship $Z_m = \mathbb{I} \{ \zeta_m > 0 \}$.  
Thus for the given data, $\bd z \defi (z_1,\ldots, z_{\Mtot})^\top$ is realized to be 1 in the first $M$ positions (and 0 for the remaining $M_0$) as a result of a latent realization $\bd\zeta \defi (\zeta_1,\ldots,\zeta_{\Mtot})^\top$ with positive values only in the first $M$ entries.

In the Bayesian framework, the introduction of the latent $\bd\zeta$ is also called \emph{data augmentation} \citep{Tanner1987} and can readily facilitate inference for parameters of certain models.  
This is indeed the case for binary data when the latent variables are normally distributed because the joint posterior of model parameters and latent variables suggests the use of Gibbs sampling \citep{Albert1993, Chib1998, DeOliveira2000}.  
This reasoning is detailed below.

By construction, the model for the observed binary data given the latent values is a point-mass
\begin{align}
   %
   [\bd Z | \bd\zeta]   
   &=  \prod_{m=1}^{\Mtot}  \mathbb{I}\{ \zeta_m >0 \}^{z_m}  +  \big( 1 - \mathbb{I}\{ \zeta_m > 0 \} \big)^{1-z_m}   \nonumber   \\
   &=   \prod_{m=1}^{M}  \mathbb{I}\{ \zeta_m >0 \}^{z_m}   \cdot   \prod_{m=1}^{M_0} \mathbb{I}\{ \zeta_{M+m} \leq 0 \}^{1-z_{M+m}}  \ .   \label{eq:LatentLik} 
\end{align}
If the latent data are modeled as depending on parameters $\bd\alpha_\zeta$, then the joint posterior for all unobservables is proportional to 
 $ [\bd Z | \bd\zeta, \bd\alpha_\zeta] \cdot [\bd\zeta , \bd\alpha_\zeta]  
=  [\bd Z | \bd\zeta] \cdot [\bd\zeta | \bd\alpha_\zeta] \cdot [\bd\alpha_\zeta] $.
Furthermore, when the latent data model $[\bd\zeta | \bd\alpha_\zeta]$ is MVN, then (\ref{eq:LatentLik}) implies that the full conditional distribution $[\bd\zeta | \bd\alpha_\zeta , \bd z]$ is a \emph{truncated} MVN and can be sampled with a Gibbs algorithm \citep{Geweke1991, Robert1995}.  
We use a model of this form because it readily admits a posterior predictive process which is also MVN and hence easy to sample from; a tractable predictive distribution is crucial to our methodology, as Section \ref{sec:GPboth} will demonstrate.

We suppose that the latent process defining the space of failures is continuous and well described by a GP-- in other words, that there is spatial correlation to the successes and failures. 
Let $\zeta_m \defi \zeta(\bd x^*_m, \bd t^*_m)$, and also suppose that $\zeta$ has mean $\mu_\zeta$ and covariance kernel $k_\zeta$ with form
\begin{align}
   k_\zeta \left( (\bd x^*_1,\bd t^*_1), (\bd x^*_2,\bd t^*_2) ; \bd\lambda_\zeta \right)   &=   \prod_{d=1}^{D_x} R \left( |x^*_{1,d}-x^*_{2,d}| ; \lambda_{\zeta,x,d} \right)   \cdot   \prod_{e=1}^{D_t} R \left( |t^*_{1,e}-t^*_{2,e}| ; \lambda_{\zeta,t,e} \right)  \ ,   \label{eq:CovLatent}
\end{align} 
where $(\bd\lambda_{\zeta,x}, \bd\lambda_{\zeta,t}) \defi \bd\lambda_\zeta$ are $D_x+D_t$ correlation length parameters.  
Note here that for identifiability, the variance of the $\zeta$ process is set to unity \citep{DeOliveira2000}.  
Therefore, in this latent model the parameter vector is $\bd\alpha_\zeta \defi (\mu_\zeta, \bd\lambda_\zeta)$, and the distribution of $\bd\zeta |\bd\alpha_\zeta$ is then MVN: 
\begin{align}
   N_{\Mtot} \Big(   
\text{mean}   =   \mu_\zeta \bd 1_{\Mtot} \ , \
\text{cov} =   k_\zeta( [\bd X^*_0 , \bd T^*_0],[\bd X^*_0 , \bd T^*_0])
   \Big)   \label{eq:MVNbin}
\end{align}
and $[\bd X^*_0 , \bd T^*_0]$ is the design matrix $[\bd X^* , \bd T^*]$ row-augmented by the experimental and calibration input settings for the runs that fail.

The posterior for all parameters in the classification problem combines the likelihood in (\ref{eq:LatentLik}), the latent prior of (\ref{eq:MVNbin}), and our choice of priors for hyperparameters: 
\begin{align}
   [\bd\zeta , \bd\alpha_\zeta | \bd z]   \ &\propto \   
   [\bd Z | \bd\zeta] \cdot [\bd\zeta | \bd\alpha_\zeta] \cdot [\bd\alpha_\zeta]   \label{eq:PostBin}  \\
   [\bd\alpha_\zeta]   &=   [\mu_\zeta] \cdot [\bd\lambda_{\zeta,x} , \bd\lambda_{\zeta,t}]   \nonumber  \\
   [\mu_\zeta]  &\propto  1   \nonumber  \\
   \lambda   &\stackrel{iid}{\sim}   \text{Unif}(0.1, 5.0)   \quad  \forall \ \lambda \in \{ \bd\lambda_{\zeta,x} , \bd\lambda_{\zeta,t} \}  \ .   \nonumber
\end{align}
Within the MCMC, the mean parameter $\mu_\zeta$ can be updated with a Gibbs step and $\bd\alpha_\zeta$ with a Metropolis step.
The elements of the latent vector are updated using the full conditional
\begin{align}
   \left[ \zeta_i | \bd\zeta_{-i} , \mu_\zeta , \bd\alpha_\zeta , \bd z \right]   
   &\propto   N \left( \mu_\zeta - \tfrac{1}{q_{ii}} \bd Q_{i,-i} (\bd\zeta_{-i} - \mu_\zeta \bd 1) \ , \   \tfrac{1}{q_{ii}} \right) 
   \cdot  \mathbb{I} \left\{ \zeta_i \in A(z_i) \right\}  \label{eq:LatentFullCond}  \\
   \bd Q  &=  \bd\Sigma^{*-1}  \quad\quad  q_{ii} = \bd Q_{i,i}  \nonumber  \\
   A(z_i)   &\defi   \left\{ \begin{array}{ll} (-\infty,0$]$ & \text{if } z_i = 0  \\  \quad (0,\infty) & \text{if } z_i = 1 \end{array} \right.  \ .   \nonumber
\end{align}
We have written the usual conditional distributions in an equivalent form featuring the precision matrix to show that $\bd\Sigma^*$ need be inverted only once \citep{Rue2005}.
All the necessary Gibbs steps can be accomplished through, e.g., the \verb1R1 package \verb1tmvtnorm1 \citep{Wilhelm2015}.
All told, the posterior sampling just described is essentially the routine of \cite{DeOliveira2000} Section 3.1, which is contained in steps $\langle 1.1 \rangle - \langle 1.2 \rangle$ of Algorithm \ref{alg:Latent} below.

The motivation behind the use of the latent GP model was the assumption that the failure mechanism is (almost surely) continuous on the latent space.
Evidence against this stipulation comes in the form of at least one very small value amongst the estimated correlation lengths, and more conspicuously, poor leave-one-out cross validation (LOOCV) classification rates. 
This may point the modeler to a specification that accounts for noise, 
such as probit GP model where the indicator functions of (\ref{eq:LatentLik}) are replaced with $\Phi(\zeta)$.
Subsequent equations would have to be adjusted accordingly.
Another option is a probit regression model \citep{Albert1993}, where the latent mean is expanded from a constant $\mu_\zeta$ to a function $\mu_\zeta(\bd x, \bd t ; \bd\beta_\zeta)$ which is linear in $\bd\beta_\zeta$, and wherein the covariance model is simplified to $iid$ $N(0, 1)$ errors.
This would greatly simplify the computations relating to (\ref{eq:PostBin}), (\ref{eq:LatentFullCond}), and (\ref{eq:PostPred}) below, but we avoided probit regression due to the undesirable model selection surrounding the choice of mean function $\mu_\zeta(\bd x, \bd t ; \bd\beta_\zeta)$.
In moderate dimensional input spaces, determining a functional basis in $(\bd x , \bd t)$ to adequately describe potential curvature in the latent process can be problematic, especially with relatively sparse data.
This was the case even in the low-dimensional illustrative example, whose pattern of failed runs was given in the right panel of Figure \ref{fig:GPcal}.

\subsection{Prediction}   \label{sec:GPbinPred}

The posterior predictive distribution for the latent process at $\tilde M$ new locations given in the rows of the matrix $[ \tilde{\bd X} , \tilde{\bd T} ]$ is denoted $\tilde{\bd\zeta}  \defi  \zeta( \tilde{\bd X}, \tilde{\bd T} )  |  (\bd\zeta , \bd\alpha_\zeta)$ and has distribution
\begin{align}
   N_{\tilde M} \Big(   
\text{mean}   &=   \mu_\zeta \bd 1_{\tilde M} + \tilde{\bd\Sigma}^\top \bd\Sigma^{*-1}(\bd\zeta - \mu_\zeta \bd 1_{\tilde M}) \ , \
\text{cov}     =   \overset{\approx}{\bd\Sigma} - \tilde{\bd\Sigma}^\top \bd\Sigma^{*-1} \tilde{\bd\Sigma}   \Big)   \label{eq:PostPred}   \\
   \bd\Sigma^*   &= k_\zeta( [\bd X^*_0 , \bd T^*_0],[\bd X^*_0 , \bd T^*_0])  \nonumber  \\
   \tilde{\bd\Sigma}  &=  k_\zeta( [\bd X^*_0 , \bd T^*_0],[\tilde{\bd X} , \tilde{\bd T}])  \nonumber \\
   \overset{\approx}{\bd\Sigma}   &=  k_\zeta( [\tilde{\bd X} , \tilde{\bd T}],[\tilde{\bd X} , \tilde{\bd T}])  \ .  \nonumber 
\end{align}
This immediately yields predictions on the binary scale: $\tilde{\bd z}{}^\top = \left(  \mathbb{I}\{ \tilde\zeta_1 >0\} , \ldots , \mathbb{I}\{ \tilde\zeta_{\tilde M} >0\}  \right)$, and this type of binary prediction can be used for predictive model assessments.
For example, as the MCMC sampling progresses, leave-one-out cross validation conditional upon the hyperparameters can be used to assess goodness-of-fit.  
The $i$th prediction is $\mathbb{I}\{ \hat\zeta_i >0\} $, where $\hat{\zeta}_i  \defi  \zeta( [\bd X_0^*, \bd T_0^*]_{i,\cdot} )  |  (\bd\zeta_{-i} , \bd\alpha_\zeta)$
is obtained from the single inversion of the full $\bd\Sigma^*$ used in the updates of the $\bd\zeta$ vector.
After the MCMC is finished, posterior summaries of the LOOCV classification rate can be calculated for the assessment of latent model adequacy.
Making posterior predictions for LOOCV can be seen as repeating step $\langle 1.3 \rangle$ of Algorithm \ref{alg:Latent} a total of $\Mtot$ times.

In addition to the predictions for cross-validation, realizations along the ``slice" $\bd t = \bd\theta$ will be of interest, and this motivates two cases.
The posterior predictive latent process can be:
\begin{enumerate}
   \item[]     
   \subitem(C1)  \quad   insensitive to $\bd x$: \ $k_\zeta \left( (\bd x_1,\bd t_1), (\bd x_2,\bd t_2) ; \bd\lambda_\zeta \right)  \equiv  k_\zeta ( \bd t_1, \bd t_2 ; \bd\lambda_{\zeta, t})$
   \subitem(C2)  \quad \hspace*{0.08in}  sensitive to $\bd x$:  \ the covariance kernel is as given above in (\ref{eq:CovLatent}).
\end{enumerate}
Case (C1) will be highly desirable because prediction along the slice amounts to a prediction at a single point.
This fact then strongly encourages model selection on the $(\bd x , \bd t)$-space: the latent GP model using the kernel of (C1) can be built first and the posterior of LOOCV classification rate can be used to assess this assumption.
Prediction at either 
$\bd\theta$ for (C1) or 
$[\tilde{\bd X} , \bd\theta]$ for (C2) 
is also a subcase of $\langle 1.3 \rangle$ in Algorithm \ref{alg:Latent}.

\begin{algorithm}[H]
\caption{Metropolis-within-Gibbs algorithm for Bayesian classification and prediction using a latent Gaussian process}   \label{alg:Latent}  
\vspace*{0.1in}
Given $\bd\alpha_\zeta^{(t-1)} = \left( \mu_\zeta^{(t-1)}, \bd\lambda_\zeta^{(t-1)} \right)$ and $\bd\zeta^{(t-1)}$  
\begin{algorithmic}
   \item[$\langle 1.1 \rangle$]  Update hyperparameters of the latent process $\bd\alpha_\zeta^{(t-1)} \rightarrow \bd\alpha_\zeta^{(t)}$ using a Gibbs step for $\mu_\zeta$ and Metropolis-Hastings update for $\bd\lambda_\zeta$ using (\ref{eq:PostBin}) 
   \item[$\langle 1.2 \rangle$]  Update latent values $\bd\zeta^{(t-1)} \rightarrow \bd\zeta^{(t)}$ with a sequence of Gibbs steps according to (\ref{eq:LatentFullCond}):
   \subitem $\bullet$  sample \ \ $ \zeta_1^{(t)}  \sim  \zeta_1  |  \left( \zeta_2^{(t-1)} , \zeta_3^{(t-1)} , \ldots, \zeta_{M_{\text{tot}}}^{(t-1)} , \ \mu_\zeta^{(t)} , \bd\alpha_\zeta^{(t)} , \bd z \right) $ 
   \subitem $\bullet$ sample \ \ $ \zeta_2^{(t)}  \sim  \zeta_2  |  \left( \zeta_1^{(t)} , \ \ \  \zeta_3^{(t-1)} , \ldots, \zeta_{M_{\text{tot}}}^{(t-1)} , \ \mu_\zeta^{(t)} , \bd\alpha_\zeta^{(t)} , \bd z \right) $
   \subitem $\vdots$
   \subitem $\bullet$  sample $ \zeta_{\Mtot}^{(t)}  \sim  \zeta_{\Mtot}  |  \left( \zeta_1^{(t)} , \zeta_2^{(t)} , \ldots, \zeta_{\Mtot-1}^{(t)} , \ \mu_\zeta^{(t)} , \bd\alpha_\zeta^{(t)} , \bd z \right) $
   \item[$\langle 1.3 \rangle$]  Sample from the posterior predictive distribution (\ref{eq:PostPred}) at $\tilde M$ locations given in the rows of some matrix $[ \tilde{\bd X} , \tilde{\bd T} ]$.
\end{algorithmic}
\end{algorithm}

Before returning to the illustrative example, we note that the modeler may be interested in better estimating the boundary of the failure-space from sequential batches of simulation runs.
Adaptively refining the estimate of the $\zeta(\bd x, \bd t) = 0$ contour under the latent GP model can be accomplished in a principled way using the Expected Improvement procedure of \cite{Ranjan2008} (pg. 530).

\subsection{Illustrative Example (cont.)}   \label{sec:Toy2}

In this example the failure mechanism is spatially correlated in $x$ and $t$ implying case (C2) and the need for two correlation length parameters $\lambda_{\zeta, x}$ and $\lambda_{\zeta, t}$ within the covariance structure:
\begin{align*}
   k_\zeta( (x_1, t_1) , (x_2, t_2) )   &=   R(x_1 - x_2 ; \lambda_{\zeta, x}) \cdot R(t_1 - t_2 ; \lambda_{\zeta, t})   \\
   R(l ; \lambda)   &=   ( 1 + \sqrt{6} | l |/\lambda ) \exp(-\sqrt{6} | l |/\lambda)  \ .
\end{align*}
The use of the Mat\'{e}rn kernel with smoothness $\nu = 1.5$ was motivated by the tendency for the covariance matrix $\bd\Sigma^*$ to become ill-conditioned.

\begin{figure}[ht]   \begin{center}
\includegraphics[width=3.0in]{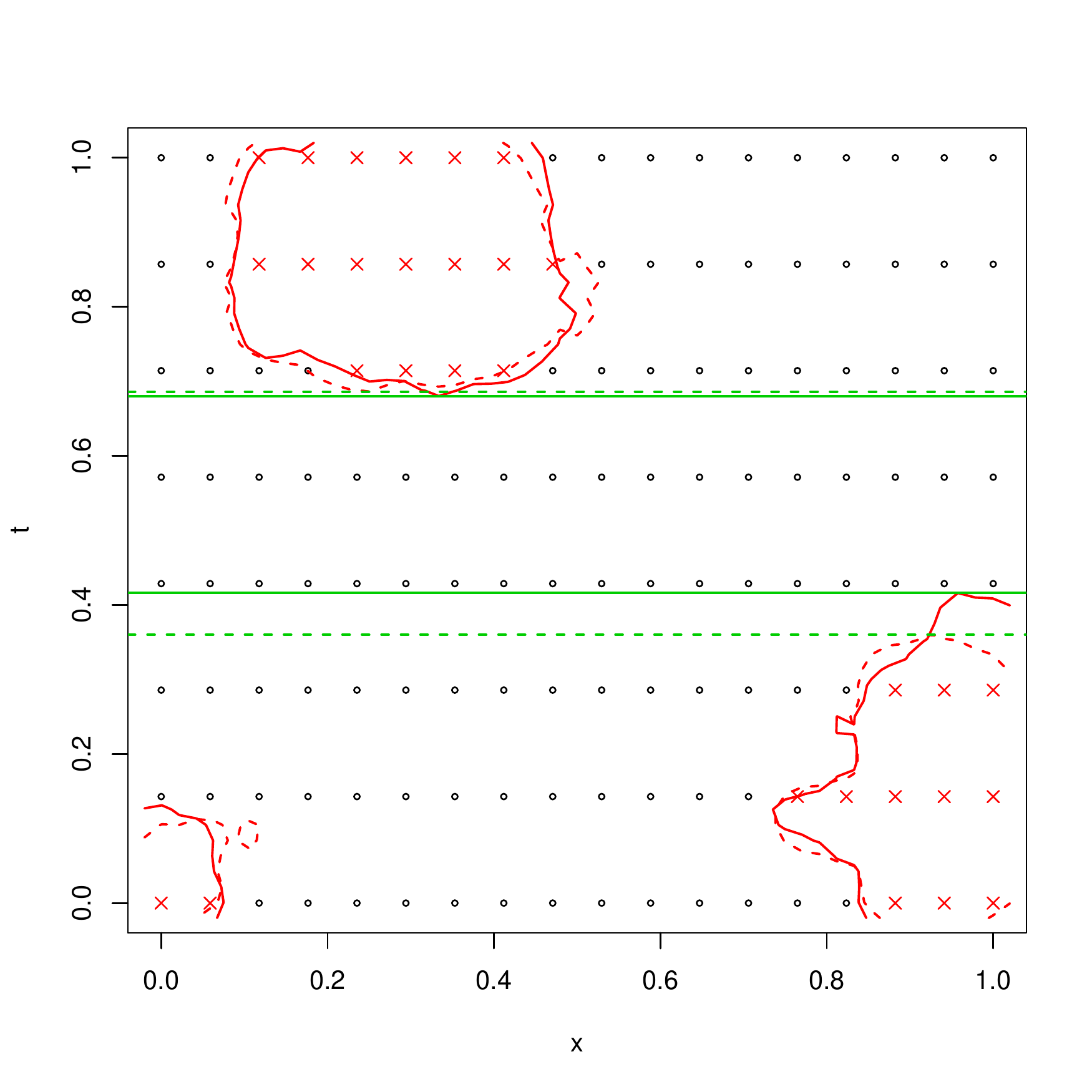}
\caption{ Contours of $\zeta = 0$ for two realizations (solid/dashed red curves) of the posterior predictive process, and corresponding ``admissible" regions whose bounds are defined by the solid/dashed green lines. }
\label{fig:GPbinPostPred}
\end{center}   \end{figure}

To fit the latent GP model, steps $\langle 1.1 \rangle$ and $\langle 1.2 \rangle$ of Algorithm \ref{alg:Latent} were performed within an MCMC of 120,000 iterations.
LOOCV predictions were performed every 200 iterations, and these resulted in a median posterior classification rate of 1.00 with a 95\% credible set of (0.993, 1.00). 
Additional predictions were made on a dense grid $[\tilde{X} , \tilde{T}]$ of $50^2$ runs, and the  $\zeta = 0$ contours for two such realizations are shown in Figure \ref{fig:GPbinPostPred}.
Each pair of green lines in the figure enclose the region for which the latent process is positive for all values of $x$.


\section{Calibration with Failed Simulator Runs}   \label{sec:GPboth}

After investigating the pattern of simulator failures and the fitting/checking the statistical model for the latent failure mechanism, the results must be discussed with the subject matter experts who provided the simulation data in order to make sense of the failed runs.
It could be, for example, that it is actually an extreme combination of variable inputs $\bd x$ (outside the range of conditions where the experimental data was obtained) is leading to output absent Property $\mathcal P$.
Or, in the case of altogether missing output, it may be that the code's numerical solvers are not converging within some configured time limit which is altogether too small.
In these two instances it would not be reasonable to exclude regions of the calibration input space.
%
Along the lines of the second instance, \cite{Huang2020} reasoned that they should allow regions where their \verb1ISOTSEAL1 code failed and that they could trust their emulator to extrapolate.
Another sensible example of ignoring computer model failures comes from the spot welding example of \cite{Bayarri2007a} (Table 3). 
Even though a high proportion of the runs failed (17/52), the absolute number is small and the failures exhibit no visible pattern as a function of their two $x$'s and one $t$; hence these seem to provide no real insight.

It is important to note that whether or not the failures are meaningful (i.e. deemed ``data"), if they depend on $\bd t$, then
two calibration analyses can be conducted regardless: one ignoring the failed runs, and one featuring the methodology of this section to utilize them.
The GP emulator within the first scenario will have to interpolate or extrapolate for regions of $\bd t$-space containing failures, potentially leading to the admission of these regions into the posterior distribution for $\bd\theta$.
This might actually be desirable in situations where the experts believe the code failures themselves are a form of discrepancy from reality.
But by comparing the posteriors from each analysis, the researchers will not only get some sense of how much the extrapolation changes the answers, but might also gain additional insight into the code.


\subsection{Model}

The basic idea of our approach is to use a carefully chosen \emph{weighted distribution} \citep{Patil1977, Bayarri1987, Patil2002} as a prior on the calibration parameters
\begin{align}
   [\bd\theta | \zeta]   &\propto   w(\bd\theta , \zeta) \cdot [\bd\theta] \ ,   \label{eq:WeightedDistrib}
\end{align}
where $w(\bd\theta , \zeta)$ is a non-negative weight function depending on the binary data $\bd z$ and additional parameters $\bd\alpha_\zeta$ via the latent function $\zeta(\bd x, \bd t)$, and $[\bd\theta]$ is the prior that would otherwise be used in a typical calibration analysis.  
This explicitly couples calibration to classification.
The approach we advocate is to use a \emph{selection model} \citep{Bayarri1987, Bayarri1992} on an unknown set that is determined by the model failures: 
\begin{align}
   w(\bd\theta , \zeta)   \defi   \mathbb{I} \left\{ \bd\theta \in \mathcal S_\zeta \right\} \ .   \label{eq:SelectionModel}
\end{align}  
In the above construction, the \emph{admissible set} $\mathcal S_\zeta$ does not (and cannot) depend on $\bd x$, but this stands to reason: if $\bd\theta$ corresponds to a true state of nature, then it should not depend on the specific configuration of experimental settings that were used/observed.  
This reasoning further suggests sets of the form
\begin{align}
   \mathcal S_\zeta   &=   \big\{ \bd\theta : \ \mathbb{P} \left( \text{success across } \bd x \text{, as determined by } \zeta \right) 
   \ \geq \   1-p_{tol}  \big\}   \label{eq:AdmissibleSet}
\end{align}
for a given tolerance probability $p_{tol}$.
A particular value of $\bd\theta$ will be said to be \emph{admissible} if it is an element of $\mathcal S_\zeta$.
When it is non-negative, the quantity $p_{tol}$ acts as a form of discrepancy. 
It acknowledges a mismatch between observed and ``true" failures in that while there may be some failed runs expected along the slice $\zeta(\bd x, \bd t = \bd\theta)$, $\bd\theta$ can still be considered admissible.  

As a concrete example, temporarily assume that the latent classifier function $\zeta(\bd x , \bd t)$ is known and deterministic. 
The set $\mathcal S_\zeta$ in (\ref{eq:AdmissibleSet}) then becomes the collection of all $\bd\theta$ such that
\begin{align}
   \mathbb{P}_{\bd\theta} \left\{ \zeta(\bd x , \bd t = \bd\theta) > 0 \right\}   &=   \int_{\mathcal X = [0,1]^{D_x}} \mathbb{I} \left\{ \bd x : \zeta(\bd x , \bd\theta) ) > 0 \right\} dG(\bd x)   \ \geq \   1-p_{tol}   \label{eq:AdmissibleSet1}
\end{align}
for choice of $p_{tol}$ and measure $G$ on the space of variable inputs. 
When
\begin{align}
   dG(\bd x) > 0 \text{ almost everywhere on } \mathcal X  \quad \text{ and } \quad  p_{tol}=0   \label{eq:AdmissCond} 
\end{align}
a particular $\bd\theta$ is admissible if and only if the minimum of the classifier function along the slice $\bd t = \bd\theta$ is positive.
The problem of checking admissibility is as such reduced to the problem of function minimization.
The reasoning extends to the case when the classifier function is unknown/random given next.

Let $\tilde\zeta(\bd x , \bd t ; \bd\alpha_\zeta)$ be a realization of a posterior predictive GP:  $\tilde\zeta  \sim  \zeta | (\bd\zeta , \bd\alpha_\zeta)$, such as the one given in (\ref{eq:PostPred}).
The conditional prior on the calibration parameters using (\ref{eq:WeightedDistrib}) and (\ref{eq:SelectionModel}), while assuming the conditions in (\ref{eq:AdmissCond}) is
\begin{align}
   [\bd\theta | \tilde\zeta]   &\propto   \cdot [\bd\theta]  \cdot  \mathbb{I} \left\{  \text{min } \tilde\zeta(\bd x , \bd t = \bd\theta ; \bd\alpha_\zeta) > 0  \right\}  \ .   \label{eq:AdmissibleSet2}   
\end{align}
Case (C1) of Section \ref{sec:GPbinPred} reduces the indicator to $ \mathbb{I} \big\{ \tilde\zeta(\bd\theta ; \bd\alpha_\zeta) > 0  \big\} $ -- i.e., evaluating the criterion at a single point.
Case (C2) will necessitate a practical compromise due to the fact that the GP cannot be realized at all points in $\mathcal X$ along $\bd t = \bd\theta$.
The simplest and most natural approximation is $ \mathbb{I} \{  \text{min } \tilde{\bd\zeta}_{\bd\theta}  > 0 \} $, where $\tilde{\bd\zeta}_{\bd\theta}  \defi \bd\zeta(\bd x = \tilde{\bd X}, \bd t = \bd\theta) | (\bd\zeta, \bd\alpha_\zeta)$ for some large space-filling design $\tilde{\bd X}$.
In either case, uncertainty in the admissible set $\mathcal S_\zeta$ will be derived from the stochastic nature of $\tilde\zeta(\bd x , \bd t ; \cdot)$ as well as from the uncertainty in the hyperparameters $\bd\alpha_\zeta$ themselves.

On a final note, we return to the discussion at the end of Section \ref{sec:GPbinMod} and admit the possibility of a probit regression model for the classifier function.
Given the regression coefficients $\bd\beta_\zeta$, a realization at the appropriate slice of the posterior predictive $\tilde{\bd\zeta}(\tilde{\bd X}, \bd\theta) | (\bd\zeta , \bd\beta_\zeta)$ is MVN with $iid$ noise and hence discontinuous everywhere.
So even though the practical check of admissibility (\ref{eq:AdmissibleSet}) on a large design $\tilde{\bd X}$ is straightforward: 
$\mathbb I \left\{ \sum_{m=1}^{\tilde M}  w_m \cdot \mathbb I \{ \tilde{\zeta}_{\bd\theta , m} > 0 \} \geq  1-p_{tol} \right\}$ 
(for pre-specified weights $w_m$ summing to unity), care must be taken to formally define the integral in the analogue of (\ref{eq:AdmissibleSet1}).
Of course, one fix is to define the admissibility criterion as using only the continuous mean of the latent process 
\begin{align}
   \int_{\mathcal X} \mathbb{I} \left\{ \bd x : \mu_\zeta(\bd x , \bd\theta ; \bd\beta_\zeta) ) > 0 \right\} dG(\bd x)   \ \geq \   1-p_{tol}  \ .   \label{eq:AdmissibleSet0}
\end{align}
This can allow for closed-form evaluations of admissibility but does not use the full posterior predictive distribution; it has stochasticity only through the regression coefficients.

\subsection{Computational Details, Discussion}

The full posterior of all the parameters is
\begin{alignat}{2}
   \big[ \bd\theta , \bd\alpha_{\eta,\delta} , \ \tilde{\bd\zeta}_{\bd\theta}, \bd\zeta , \bd\alpha_\zeta \ | \ \bd d , \bd z \big]
   &=   \big[ \bd\theta , \bd\alpha_{\eta,\delta} , \ \tilde{\bd\zeta}_{\bd\theta} \ | \ \bd\zeta , \bd\alpha_\zeta , \ \bd d , \bd z \big]   
   &&  \cdot   \big[ \bd\zeta , \bd\alpha_\zeta \ | \ \bd d , \bd z \big]   \nonumber  \\
   &=   \big[ \bd\theta , \bd\alpha_{\eta,\delta} \ | \ \tilde{\bd\zeta}_{\bd\theta} , \ \bd d \big]   \cdot   
   \big[ \tilde{\bd\zeta}_{\bd\theta} \ | \ \bd\zeta , \bd\alpha_\zeta \big] 
   &&  \cdot   \big[ \bd\zeta , \bd\alpha_\zeta \ | \ \bd z \big]   \label{eq:FullPosterior}
\end{alignat}

\begin{itemize}
   \item   The first term in (\ref{eq:FullPosterior}) is proportional to $L \left(\bd\theta , \bd\alpha_{\eta,\delta} ; \ \bd d \right)  \cdot  [ \bd\theta \ | \ \tilde{\bd\zeta}_{\bd\theta} ]  \cdot  [ \bd\alpha_{\eta,\delta} ]$ which is the posterior for a traditional calibration model, except that a marginal prior for $\bd\theta$ is replaced by a conditional prior $[ \bd\theta \ | \ \tilde{\bd\zeta}_{\bd\theta} ]$ such as that of (\ref{eq:AdmissibleSet2}).
   \item   The second term is the posterior predictive distribution of the latent process (\ref{eq:PostPred}) along the $\bd t = \bd\theta$ slice.
   \item   The third term is the posterior for the latent process values and hyperparameters of the classification model (\ref{eq:PostBin}).
   Here it is explicit that $\zeta$ is assumed independent of $\eta$ and $\delta$ (further, the response values of the successful simulations and experimental data).
\end{itemize}
This clean factorization allows for a straightforward implementation using the results of previous sections.
The pseudocode of the whole procedure is given in Algorithm \ref{alg:CalibWithFail}.

\begin{algorithm}[H]
\caption{Metropolis-within-Gibbs algorithm for Bayesian calibration with simulator failures}   \label{alg:CalibWithFail}
\begin{algorithmic}
\vspace*{0.1in}
   \item[$\langle 2.1 \rangle$]  Update latent values and hyperparameters as in $\langle 1.1 \rangle$ and $\langle 1.2 \rangle$ resulting in $\bd\zeta^{(t)}$ and $\bd\alpha_\zeta^{(t)}  =  (\mu_\zeta, \bd\lambda_\zeta)^{(t)}$  
   \item[$\langle 2.2 \rangle$]  Update emulator and discrepancy parameters using (\ref{eq:PostCal}) and the description within Section \ref{sec:GPcal} resulting in $\bd\alpha_{\eta , \delta}^{(t)}  =  (\mu_\eta, \mu_\delta, \ \sigma^2_\eta, \sigma^2_\delta, \sigma^2_\epsilon, \  \bd\lambda_{\eta,x}, \bd\lambda_{\eta,t}, \bd\lambda_\delta)^{(t)}$ 
   \item[$\langle 2.3 \rangle$]  Update calibration parameters $\bd\theta$ with a Metropolis step: 
      \subitem $\bullet$   
      propose $\bd\theta^*$ from distribution having density $q( \bd\theta^\text{new} | \bd\theta^\text{old} = \bd\theta^{(t-1)} )$ 
      \subitem $\bullet$
      check admissibility using the indicator $ \mathbb{I} \{  \text{min } \tilde{\bd\zeta}_{\bd\theta^*}  > 0 \} $ from predictions $\langle 1.3 \rangle$ according to one of the following cases
         \subsubitem \hspace*{10pt} (C1)  \quad  $\tilde{\bd\zeta}_{\bd\theta^*}  \sim  \zeta(\bd t = \bd\theta^*)| (\bd\zeta^{(t)} , \bd\alpha_\zeta^{(t)})$  
         \subsubitem \hspace*{10pt} (C2)  \quad  $\tilde{\bd\zeta}_{\bd\theta^*}  \sim  \zeta(\bd x = \tilde{\bd X}, \bd t = \bd\theta^*) | (\bd\zeta^{(t)} , \bd\alpha_\zeta^{(t)})$ for some large space-filling design $\tilde{\bd X}$
      %
      \subitem $\bullet$
      conditional upon admissibility, set $\bd\theta^{(t)} = \bd\theta^*$ with probability min$(1,R)$ where 
      $$R = \frac{ L(\bd\theta^* \ ; \ \bd\alpha_{\eta , \delta}^{(t)} , \bd d) \cdot [\bd\theta^*] \cdot q( \bd\theta^{(t-1)} | \bd\theta^* ) }{ L(\bd\theta^{(t-1)} \ ; \ \bd\alpha_{\eta , \delta}^{(t)} , \bd d) \cdot [\bd\theta^{(t-1)}] \cdot q( \bd\theta^* | \bd\theta^{(t-1)} ) } $$
\end{algorithmic}
\end{algorithm}

It is very important to observe that the classification and calibration models can be investigated independently before the coupled analysis.
The latent failure process and its hyperparameters are independent of the output values of successful simulation runs, so the classification model should be analyzed first to get some idea how useful the observed failures will be.
Starting values and proposal distributions for the classification analysis can be used directly within the MCMC of the coupled model.
The calibration model without the use of failed simulator runs can also be fit independently in order to get a reasonable initial proposal distributions for the combined analysis.
Fitting both models separately will also allow the researcher to determine whether hyperparameters of either can be fixed, if desired.

How useful are the failed runs?
Insight is provided by first considering the case of a fixed, known constraint.
For clarity, and without loss of generality, one may further suppose the emulator and discrepancy hyperparameters within $\bd\alpha_{\eta , \delta}$ are known.
Property $\mathcal P$ is guaranteed when the full domain of $\bd\theta$ is restricted to a subset $\mathcal S$ (temporarily suppressing the ``$\zeta$" subscript)
implying the constrained posterior is 
\begin{align}
  [\bd\theta \ | \ \bd d, \mathcal P]  
  \hspace*{5pt} \propto \hspace*{5pt}
%
  L(\bd\theta ; \bd d) \cdot [\bd\theta] \cdot \mathbb{I} \left\{ \bd\theta \in \mathcal S \right\}  
  \hspace*{5pt} \propto \hspace*{5pt}
  [\bd\theta \ | \ \bd d]  \cdot \mathbb{I} \left\{ \bd\theta \in \mathcal S \right\}  \ .  \label{eq:ConstrainedPosterior}  
\end{align}
Furthermore, the normalizing constant of the constrained posterior,
$ \pi \defi \int_{\mathcal S} [\bd\theta | \bd d] d\bd\theta $,
is obviously $\mathbb P \left\{ \bd\theta \in \mathcal S \right\}$ under the full posterior.
If this probability is close to 1 then the restriction to $\mathcal S$ is not actually very restrictive.
If given a sample $\bd\theta^{(1)},\ldots, \bd\theta^{(N_{tot})} \stackrel{iid}{\sim} [\bd\theta | \bd d]$, then (\ref{eq:ConstrainedPosterior}) implies that 
$\{\bd\theta^{(n)} : \bd\theta^{(n)} \in \mathcal S\}$
is a sample from the constrained posterior.
An estimate 
of its normalizing constant is $\widehat\pi = \frac{1}{N_{tot}} \sum_{n=1}^{N_{tot}} \mathbb I \{ \bd\theta^{(n)} \in \mathcal S \}$.

In the case of a random constraint, uncertainty in the hyperparameters $\bd\alpha_\zeta$ and predictive process $\tilde\zeta | (\bd\zeta, \bd\alpha_\zeta)$ for the latent classifier can be factored into an assessment of how useful the failed runs are.
Consider the binary matrix $\bd B$ whose columns correspond to draws of the calibration parameters under the traditional posterior (disregarding the failures) and whose rows correspond to draws of the latent posterior $\big[ \tilde{\bd\zeta}, \bd\zeta, \bd\alpha_\zeta  | \bd z \big]$;
the $(i,j)$ entry of $\bd B$ is 1 if $\bd\theta^{(j)}$ is predicted as a success for realization $i$ of the latent parameters and 0 otherwise.
The row-means of this matrix provide a distribution of $\pi$ estimates as the latent quantities vary.
If there is relatively little uncertainty in this distribution and it is very close to 1, the researcher need not conduct the full coupled sampling described in Algorithm 2--
an approximate sample from the constrained posterior is obtained by simply omitting values from the full sample that are not in the admissible set $\mathcal S_{\widehat\zeta}$ derived from $\mathbb E\{ \tilde\zeta \ | \ \bd\zeta, \widehat{\bd\alpha}_\zeta \}$ and $\widehat{\bd\alpha}_\zeta = \mathbb E\{ \bd\alpha_\zeta \}$.

The column-means of the binary prediction matrix $\bd B$ are arguably more informative.
The mean of $\bd B_{\cdot,j}$ is the estimated posterior probability $\widehat\pi_j$ of $\bd\theta^{(j)}$ being admissible (marginalizing over the latent quantities).
This collection of values give a distribution on the pointwise probability of admissibility.
The proportion of $\{ \widehat\pi_j \}_{j=1}^{N_{tot}}$ which are approximately zero indicate the fraction of
$\bd\theta^{(1)},\ldots \bd\theta^{(N_{tot})}$ which are almost always classified as failures-- values for which the failed runs would be considered informative.
On the other hand the proportion which are approximately unity indicate the fraction of full posterior values which would be admissible with or without the use of the failed simulation runs.
For the analysts not willing to incorporate the admissibility criterion within their calibration routine, 
the collection $\{ \widehat\pi_j \}_{j=1}^{N_{tot}}$ can be used as resampling weights for the traditional posterior $\bd\theta$ draws.
If estimation is not the sole purpose of calibration, then the full coupling of Algorithm \ref{alg:CalibWithFail} should be used, as the updates for the discrepancy parameters (vital for prediction) will be affected by the restricted $\bd\theta$ updates.



\section{Examples}   \label{sec:Examples}

Here we demonstrate the methodology with the illustrative example as well as the problem that originally motivated our calibration technique.

All proposal distributions for Metropolis steps were taken to be MVN after parameter transformations.
For example, the univariate marginal priors on the calibration parameters were first rescaled to [0,1] and then to $(-\infty, \infty)$ via a probit transformation $\Phi^{-1}(\cdot)$.
All such MVN proposals had their covariances tuned adaptively according to \cite{Haario1999, Haario2001}.

\subsection{Illustrative Example (cont.)}

Failures were observed at $t < 0.286$ and $t > 0.714$ (that is, for at least one value of $x$) suggesting that that the true set of admissible $\theta$ is contained in the interval between these two values.
As we shall see, the calibration analysis incorporating failed runs will indeed produce a posterior distribution for $\theta$ having support $\approx (0.286, 0.714)$, though with tails weighted according to the uncertainty in the latent failure process.

After fitting and checking the latent model, 
the coupled analysis was conducted;
Algorithm \ref{alg:CalibWithFail} was used to obtain 120,000 draws of all parameters.
The admissibility of a proposed $\theta^*$ was checked using a draw of the latent posterior predictive process on the slice with $\tilde X$ containing 50 equally spaced points.
Had the entire latent process been realized, the $\zeta = 0$ contours would appear akin to the red curves in Figure \ref{fig:GPbinPostPred}.
Each set of red curves would define the boundaries (green lines) of the admissible set, and a proposed $\theta^*$ would be admissible if it fell within these boundaries.
There is a small probability that a realization of the latent process is positive within the naive bounds $(0.286, 0.714)$, and for that instance the admissible set is a union of intervals; this probability is smaller for longer correlation lengths and higher smoothness of the latent GP.

\begin{figure}[h!t]   \begin{center}
\includegraphics[width=3.0in]{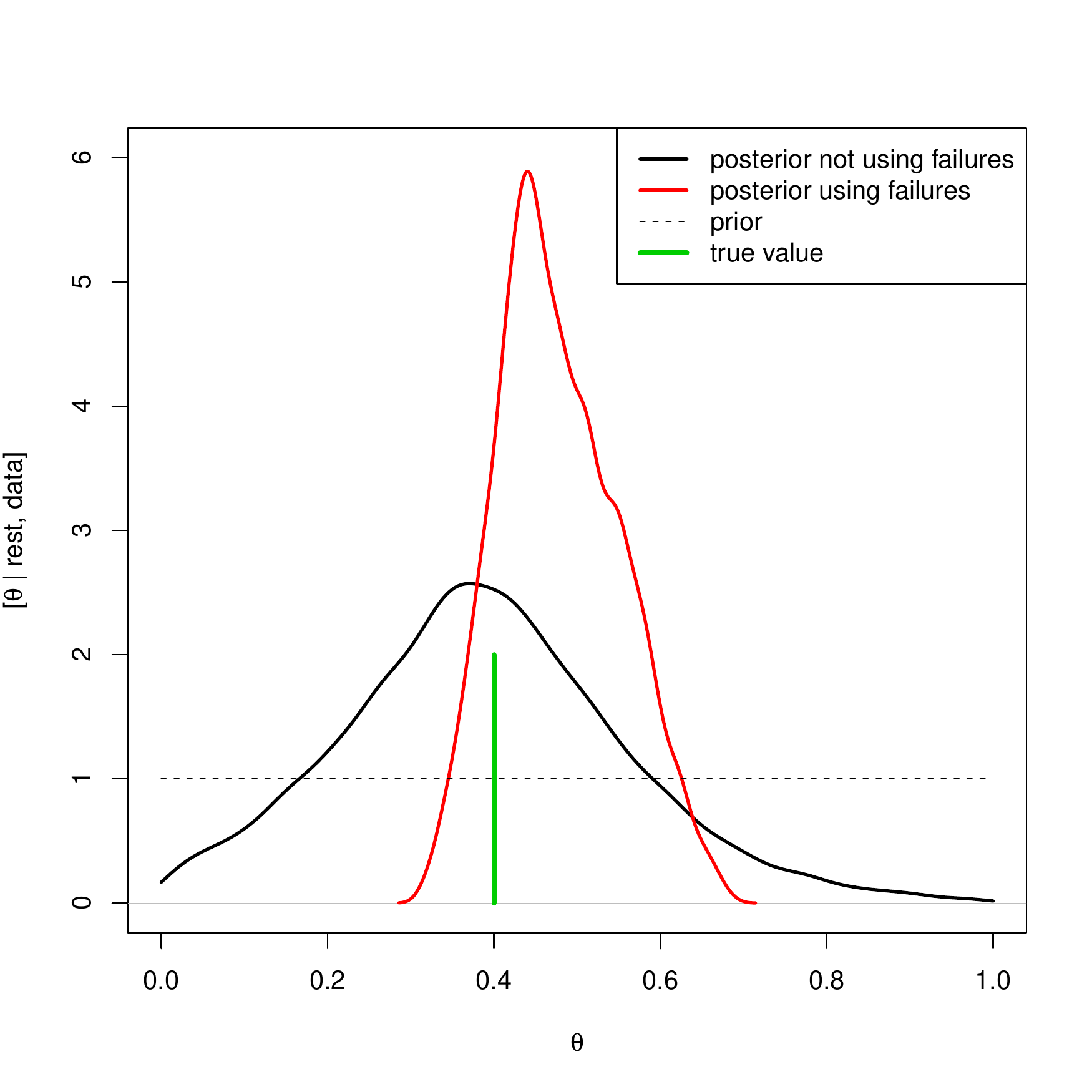}
\caption{ Posterior distributions for calibration parameters with and without using the failed simulations. }
\label{fig:PosteriorCompare01}
\end{center}   \end{figure}

The posterior distribution of the calibration parameters without and with using the failures is displayed in Figure \ref{fig:PosteriorCompare01}.
In the latter case, there is zero probability given to regions that fail for some $x$, as desired.
However the distribution incorporating failed runs is not the full posterior restricted to the naive bounds because there is uncertainty associated with these bounds.
In fact, the probability of admissibility (under the posterior for the latent parameters) ranges from about 0.22 to 0.61;
to contrast, the point estimate of this value using the naive bounds is nearly 0.7.
Regarding pointwise admissibility, 37\% of the samples from the full posterior were always classified as failures whereas 20\% were always predicted to be successes.

For this toy problem the true calibration parameter and traditional posterior mean/ mode/ median fell squarely within the admissible region, but this need not be.
In fact, were the inclusion of failures to rule out a region having high probability under the usual calibration case, this might challenge the researchers' understanding of the numerical model or physical system.
It would also highlight the fact the most probable configurations of parameters were inferred by extrapolation of the response surface to regions which had no successful runs.
As such this would be an interesting and important scenario.

\subsection{MFIX CFD model}

We conclude this paper by applying the methodology to the example that motivated it.

\begin{table}[htb]   \scriptsize
\centering
\caption{Summary of inputs, outputs, and CFD calibration parameters.  Here, trNorm$(\mu, \sigma^2; [a,b])$ is a N$(\mu, \sigma^2)$ distribution truncated to the interval $[a,b]$, and ssBeta$(\alpha, \beta; [a,b])$ is a Beta$(\alpha, \beta)$ distribution shifted and scaled to the interval $[a,b]$. \vspace*{8pt}}
\label{tab:InputOutput}
\begin{tabular}{| l l |}
\hline 
\hline
\underline{ Output \ $\bd y$: }   &  \\
\quad $y$: knot 0.975 of the functional breakthrough curve   &  \\   [2ex]
\underline{ Experimental Conditions (Variable Inputs) \ $\bd x$: }   &   \underline{ Range/Prior }  \\   [1ex]
\quad $x_1$: gas inflow rate  (std. L/min)   &   [15.0, 30.0]    \\
\quad $x_2$: partial pressure of CO$_2$  (\%)   &   [10.0, 20.0]  \\
\quad $x_3$: coil temperature  ($^\circ$C)   &   [39.0, 81.5]   \\
\quad $x_4$: gas inflow temperature  ($^\circ$C)   &   [23.6, 38.4]   \\   [1ex]
\underline{ CFD Model Parameters (Calibration Inputs) \ $\bd t$: }   &  \\
dry adsorption  &  \\
\quad \  $t_1$:  formation enthalpy (J/mol)  \hspace*{0.55in}  $\Delta H_c$   &   trNorm$\left( -7.8\ttplus 4, (1.18\ttplus 4)^2; \ [-1.5\ttplus 5, -3.0\ttplus 4] \right)$  \\
\quad \  $t_2$:  formation entropy  (J/mol$\cdot$K)  \hspace*{0.48in}  $\Delta S_c$   &   trNorm$\left( -250, 25^2; \ [-320,-200] \right)$  \\
\quad \  $t_3$:  activation energy  (J/mol)  \hspace*{0.65in} $\Delta H_c^\ddag$
     &   Unif$(3.0\ttplus 4, 1.2\ttplus 5)$  \\
\quad \  $t_4$: log pre-exponential factor  (unitless)  \hspace*{0.13in}  $\log_{10} C_c$
     &   Unif$(0, 5.5)$  \\
water physisorption  &  \\
\quad \  $t_5$: formation enthalpy  \hspace*{0.99in}  $\Delta H_a$   &   Unif$(-1.0\ttplus 5, -1.0\ttplus4)$  \\
\quad \  $t_6$: formation entropy  \hspace*{1.04in}  $\Delta S_a$   &   Unif$(-320, -200)$  \\
\quad \  $t_7$: activation energy  \hspace*{1.09in}  $\Delta H_a^\ddag$
     &   Unif$(2.0\ttplus 4, 1.2\ttplus 5)$  \\
\quad \  $t_8$: log pre-exponential factor  \hspace*{0.65in}  $\log_{10}C_a$   &   Unif$(0, 4)$  \\
wet reaction  &  \\
\quad \  $t_9$: formation enthalpy  \hspace*{1.00in}  $\Delta H_b$   &   Unif$(-1.5\ttplus 5, -3.0\ttplus 4)$  \\
\quad $t_{10}$: formation entropy  \hspace*{1.04in}  $\Delta S_b$   &   Unif$(-320, -200)$  \\
\quad $t_{11}$: activation energy  \hspace*{1.09in}  $\Delta H_b^\ddag$   &   Unif$(2.5\ttplus 4, 1.2\ttplus 5)$  \\
\quad $t_{12}$: log pre-exponential factor  \hspace*{0.65in}  $\log_{10}C_b$   &   Unif$(0, 4)$ \\   [2ex]
\quad $t_{13}$: particle size ($\mu$m)   &   ssBeta(4.5, 3.3; 108, 125)      \\
\quad $t_{14}$: effective amine proportion when fresh (unitless)   &   trNorm$\left( 0.177, \ (0.027)^2; \ [0.133, 0.210] \right)$  \\ 
\hline
\end{tabular}   \normalsize
\end{table}

The Carbon Capture Simulation Initiative (CCSI, a partnership between Department of Energy laboratories, academic institutions, and industry) has been developing state-of-the-art computational tools to assist in the development of technologies that capture CO$_2$ from coal-burning power plants.  
One technology investigated was an adsorber system with a bubbling fluidized bed of 32D1 sorbent \citep{Lane2014, Spenik2015}.  
In this system, a mixture of gases that includes CO$_2$ flows up through the bed of solid sorbent particles and together they form a fluid-like state. 
The increased mixing between gases and solids better facilitates the adsorption of CO$_2$ from the gas stream, that is, it increases the adhesion of the CO$_2$ atoms to the surface of the sorbent particles.

CCSI's investigation of this carbon capture apparatus was done sequentially using a validation hierarchy \citep{Lai2016}.  
Experiments of increasing complexity were formulated to isolate subsystems of the entire multiphase flow process; computer models of each idealized system were also formulated such that each could be calibrated to its corresponding physical experiment.  
In particular we use data from just one of these validation exercises:  the case of hot-reacting flow [Sec. 6.3 of \cite{Lai2016}].

The experimental data were obtained by varying four quantities: gas inflow rate, partial pressure of CO$_2$, coil temperature, and gas inflow temperature.
A Latin Hypercube design of size 52, with 19 points replicated (for a total of $N = 71$ points) was used to cover the variable input space; the bounds of 4-dimensional cube can be found in Table \ref{tab:InputOutput}). 
Each experiment produced a breakthrough curve describing the CO$_2$ adsorption over time, and this monotonic functional output could very accurately be captured by five points: the $y$-intercept and the time until 2.5, 25, 50, and 75 percent reductions from the intercept (initial adsorption) via a monotone cubic log-spline.
The second of these five outputs (knot 0.975) contained a large percent of the total variation, so we restrict our analysis to this single quantity.

The numerical representation of the experimental system was a computational fluid dynamics (CFD) model implemented within the 
MFIX (Multiphase Flow with Interphase eXchanges) software \citep{Syamlal1993}.
MFIX solves the governing equations for conservations of mass, momentum, energy, and species subject to the boundary and initial conditions in multiphase flow.

The physics of 32D1 reactive flow involve multiple mechanisms including the hydrodynamics of multiphase flow, the transfer of heat, and the reactions between chemical species within the mixed system of the fluidized bed.
Three main chemical reactions occur when 32D1 adsorbs carbon dioxide: 
the reaction of CO$_2$ with the impregnated amine to form carbamate (dry adsorption, \ref{eq:DiffEq1}), 
the physical adsorption of H$_2$O to the sorbent (water ``physisorption", \ref{eq:DiffEq2}), 
and
the reaction of CO$_2$, physisorbed H$_2$O, and amine to form bicarbonate (wet reaction, \ref{eq:DiffEq3}).
The site fractions of carbamate anion, adsorbed water, and bicarbonate are denoted $c$, $a$, and $b$, respectively.   
\begin{align}
   \first{c}{t}   &=   C_c  \ T \exp \left( \frac{-\Delta H_c^\ddag}{RT} \right)  \cdot  f_c \left( c , b ; \ \Delta S_c , \Delta H_c \right)  \label{eq:DiffEq1}  \\
   \first{a}{t}   &=   C_a  \ T \exp \left( \frac{-\Delta H_a^\ddag}{RT} \right)  \cdot  f_a \left( a ; \ \Delta S_a , \Delta H_a \right)  \label{eq:DiffEq2}  \\
   \first{b}{t}   &=   C_b  \ T \exp \left( \frac{-\Delta H_b^\ddag}{RT} \right)  \cdot  f_b \left( c , a , b ; \ \Delta S_b , \Delta H_b \right)    \label{eq:DiffEq3}   
\end{align}   
Above, $T$ is the temperature of the reacting system and $R$ is the gas constant.   
Each ordinary differential equation within the coupled system depends on four parameters, $\Delta H, \Delta S, \Delta H^\ddag, C$ and the subscripts ``$c$", ``$a$", and ``$b$" indicate the appropriate reaction (\ref{eq:DiffEq1}, \ref{eq:DiffEq2}, \ref{eq:DiffEq3}).
The $f_c$, $f_a$, and $f_b$ functions within the rate equations have relatively simple forms but are excluded because of the additional definitions required.
Two additional parameters were included for calibration, one related to effective particle size of the sorbent and one to pertaining to amine degradation.
The expanded formulation of the physics and a more detailed account of the parameters can be found in \cite{Lai2016}.   

\begin{figure}[h!t]   \begin{center}
\includegraphics[width=4.0in]{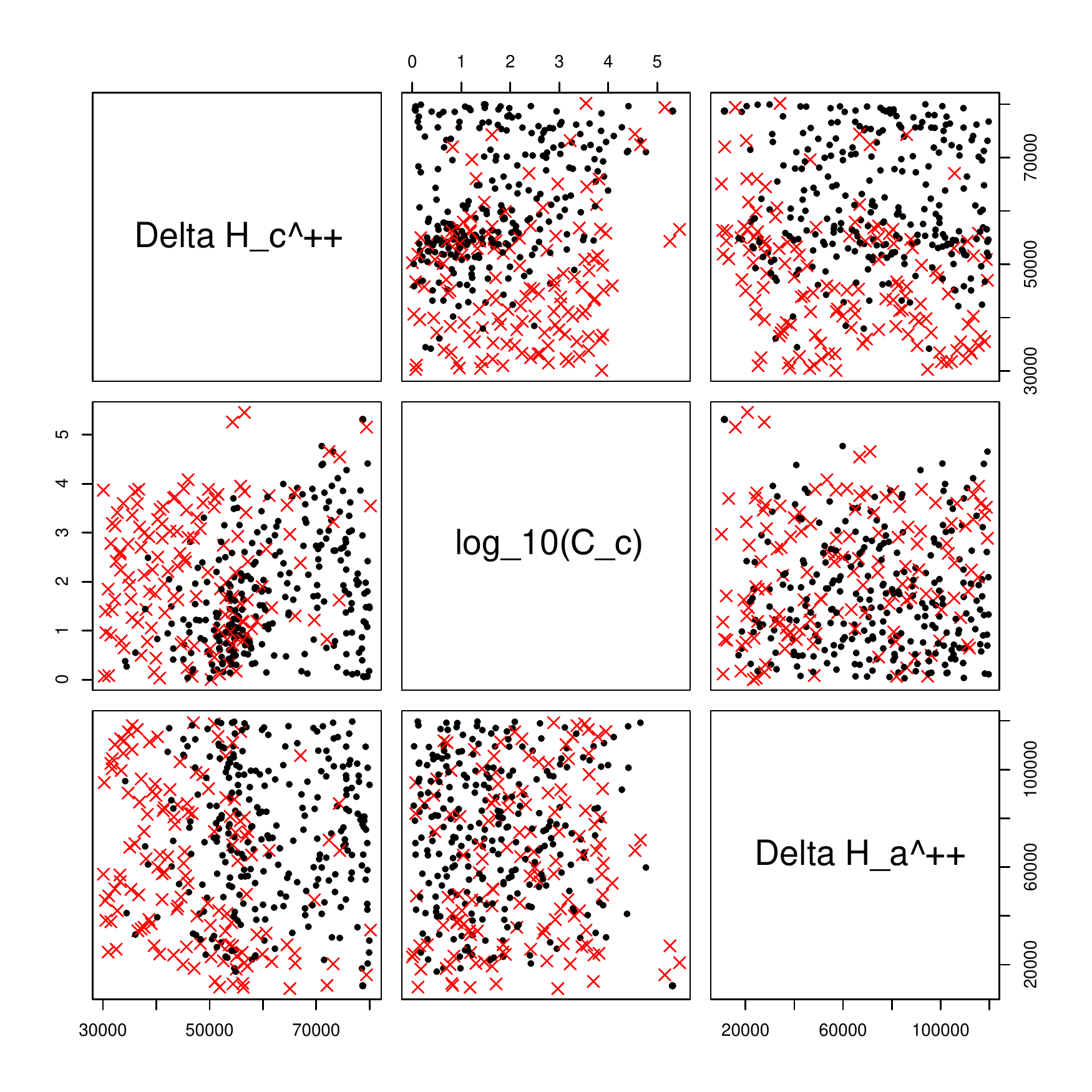}
\caption{ The successes/failures (black dots / red $\times$'s) versus three calibration inputs of the MFIX CFD code. }
\label{fig:Failures}
\end{center}   \end{figure}

A total of $\Mtot = 471$ simulations were run, but of these $M_0 = 136$ failed to converge (implying $M = 335$).  
With this large percentage of failed runs the researchers wanted to utilize this information, if possible.
This was largely due to necessity as subsequent uncertainty analyses would also be based around MFIX-- a code whose numerical solvers are not easily modified.
In general, just because a code fails to produce output it does not imply that there is not \emph{some} numerical regime which could produce reasonable answers at the given input settings.
This is why the simulation runs must be discussed with the subject-matter experts; further, it should be decided whether or not calibration should be allowed to include regions where the emulator is forced to extrapolate.

The chemists had indicated that extreme reaction rates would be numerically problematic (and likely to produce unrealistic output even if convergent),
but could not provide an explicit relationship between calibration parameter configurations and success/failure.
Though the bounds within the prior distributions (Table \ref{tab:InputOutput}) were a good faith effort to encapsulate previous knowledge of the physics and failure-space boundaries, they could not completely preclude troublesome combinations.
Indeed, a number of projection-based exploratory plots of the failures were studied but clear patterns did not emerge, some evidence that three or more of the parameters were jointly causing problems.
An exception is found in Figure \ref{fig:Failures} where it can be seen that model failures frequently result from low $\Delta H_c^\ddag$ and high $\log_{10}C_c$ 
(i.e. when the dry adsorption occurs too quickly, in a relative sense).
As for the low values of $\Delta H_a^\ddag$ producing instabilities, this was not known \emph{a priori} but it was concluded that the lower bound for water physisorption activation energy was likely too pessimistic.


The latent Bayesian model under case (C1) (no dependence of the failures upon the experimental conditions) and Squared-Exponential latent correlation was run for 200,000 iterations, and after a short burn-in, the chains showed no evidence of non-convergence.
Every 200 iterations LOOCV was performed resulting in a classification rate between 80.0\% and 98.3\% with a median of 91.3\%. 
Given the dimension of calibration input space, these rates were deemed adequate enough to proceed with the coupled calibration under (C1).

\begin{figure}[h!t]   \begin{center}
\includegraphics[width=4.5in]{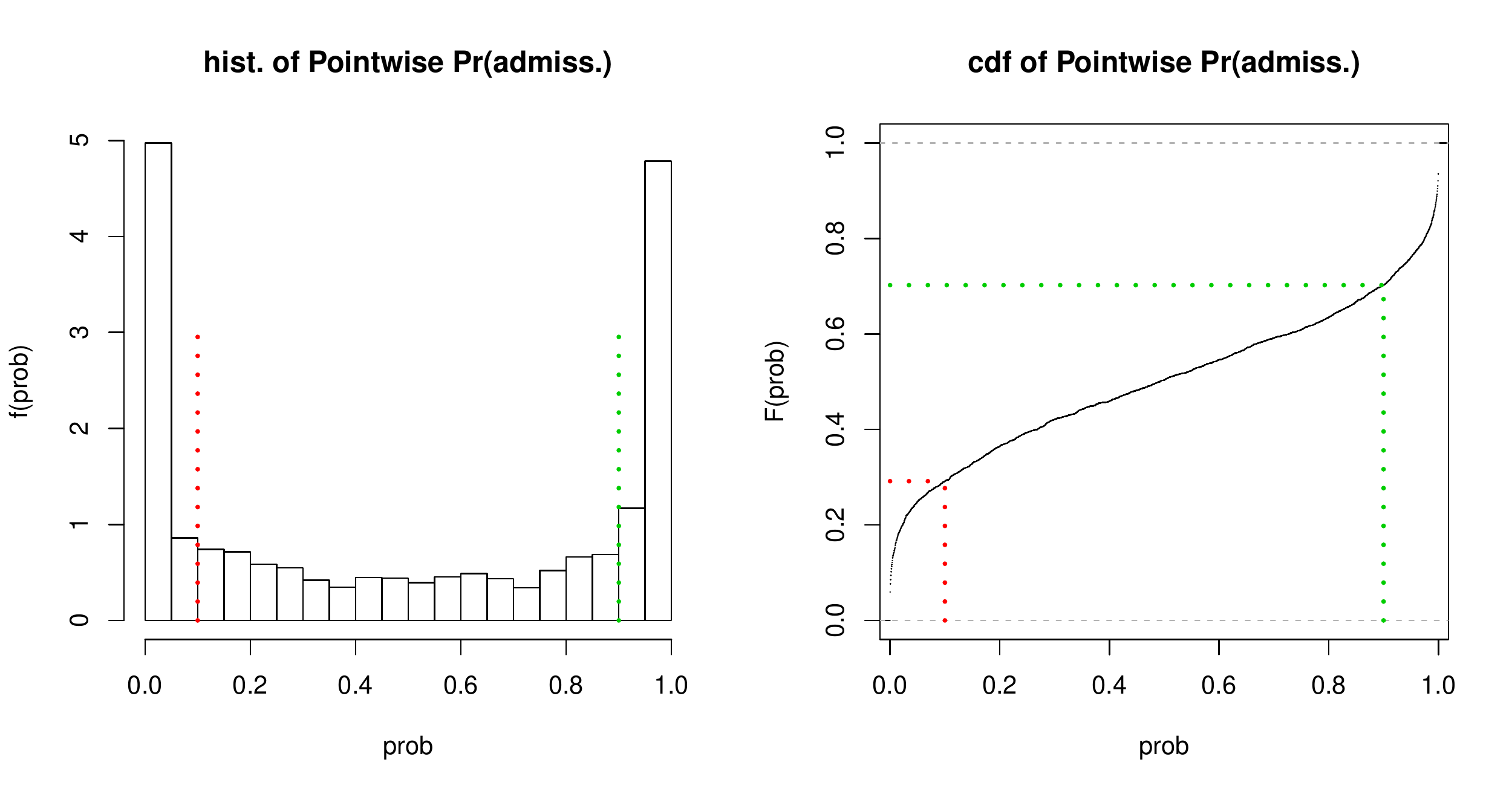}
\caption{ Estimates of pdf and cdf for the pointwise probability of admissibility of the full posterior calibration parameter samples under the uncertain classifier.  The dashed line segments indicate the lower and upper 10\%-- i.e., proportions of the full posterior samples which are almost always classified as failures and successes, respectively. }
\label{fig:ProbAdmiss}
\end{center}   \end{figure}

Calibrations were conducted without and with the failed simulations.
(It should be noted that the results of the analysis without using the failed runs differ from those of \cite{Lai2016} because different models and data were used-- the current work used only one of the five outputs and a GP model instead of a Bayesian Smoothing Spline ANOVA model.)
The priors for the 14 calibration parameters are given in Table \ref{tab:InputOutput}. 
They are simple univariate uniform, truncated normal, or shifted/scaled beta distributions obtained from a previous analysis.
For each case three chains of 200,000 iterations were used, recording every third sample after a burn-in of 50,000.
Binary predictions of success/failure were obtained using a subset of draws from the traditional calibration posterior (ignoring the failed runs) and classifiers based upon the posterior for the latent quantities.
The estimated distribution of pointwise admissibility probabilities is shown in Figure \ref{fig:ProbAdmiss}.
In particular, while 29.1\% of the posterior samples were almost always classified as successes, 28.3\% of the calibration parameter samples were almost always ruled out as inadmissible.
This suggested that the failures were indeed informative and that the coupled calibration could be performed in order to properly weight $\bd\theta$-space according to uncertainty in the latent classifier.

\begin{figure}[h!t]   \begin{center}
\includegraphics[width=3.0in]{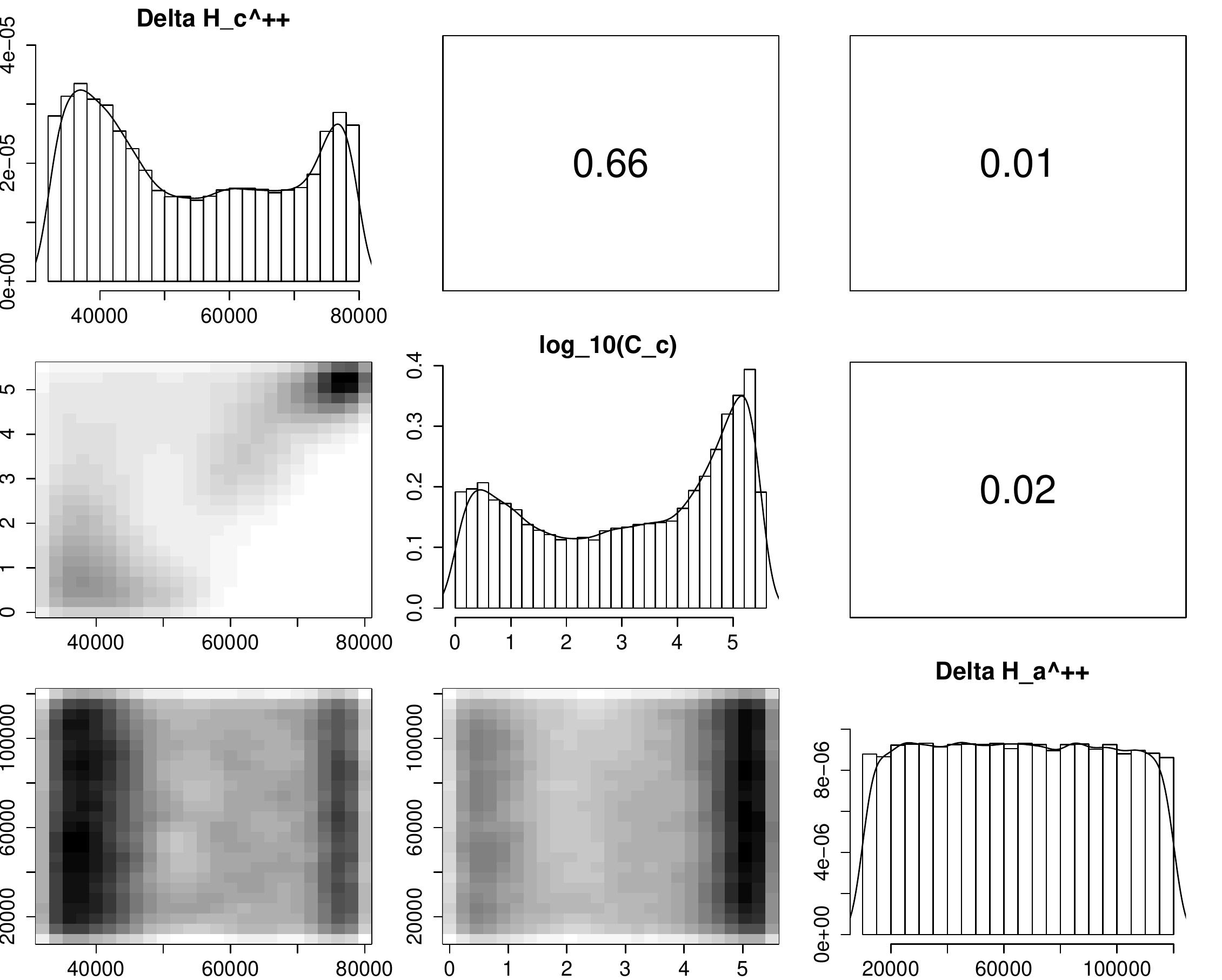}
\hspace*{0.2in}
\includegraphics[width=3.0in]{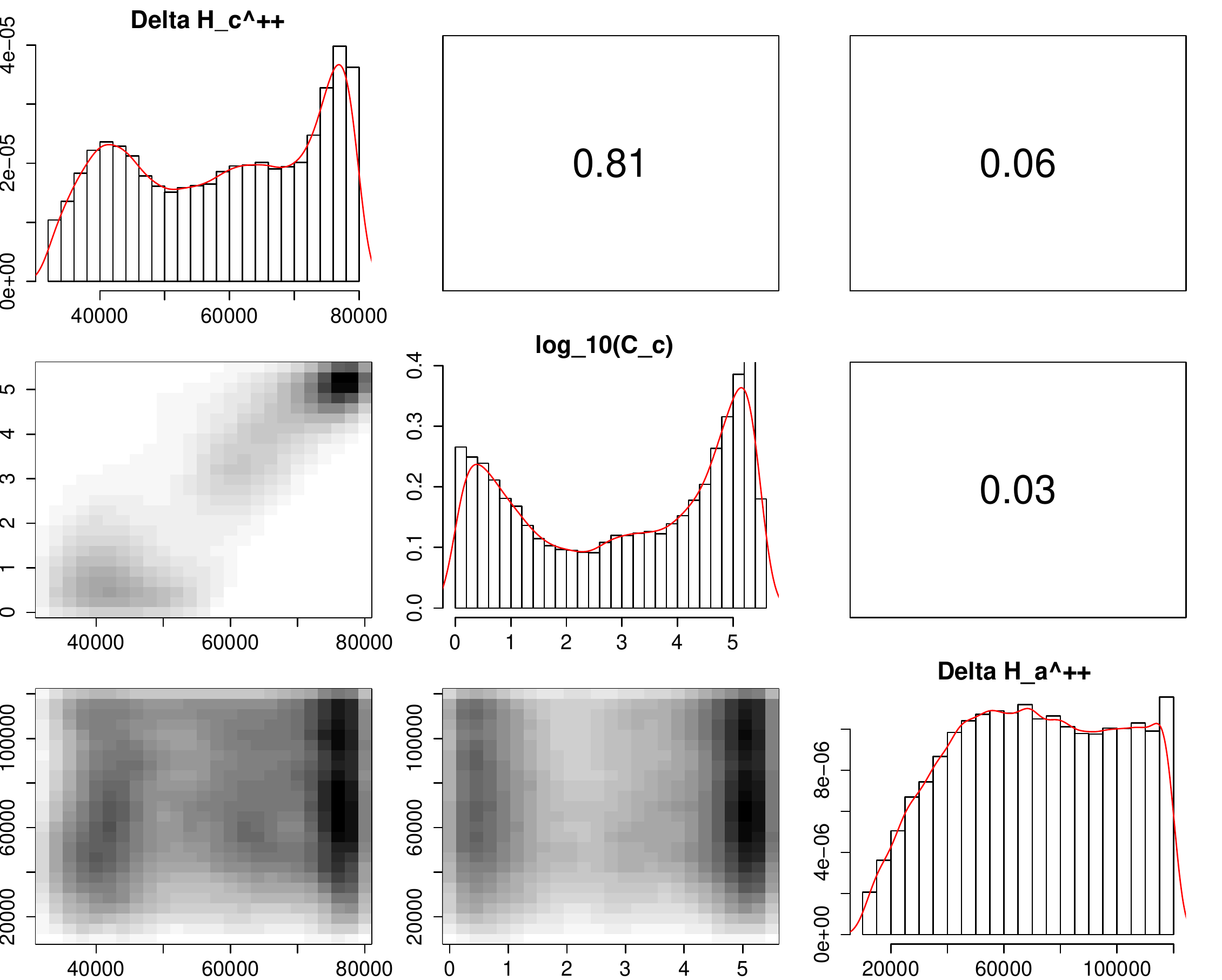}
\caption{ Posterior distributions for three calibration parameters without and with using failed runs (left/right). The numbers in the panels above the diagonal are estimated pairwise correlations. }
\label{fig:PosteriorCompare14}
\end{center}   \end{figure}

The bivariate marginal posterior distributions of the calibration parameters were compared and it was seen that 11 of the 14 parameters had density estimates which appeared qualitatively similar.   
The three parameters that differed most were  
$\Delta H_c^\ddag$, $\log_{10}C_c$, and $\Delta H_a^\ddag$; 
the posteriors for each case are given in Figure \ref{fig:PosteriorCompare14}.
As expected, the posterior utilizing the failed runs excluded regions with 1) low $\Delta H_c^\ddag$, high $\log_{10}C_c$, and 2) low $\Delta H_a^\ddag$.
The fact that the latent LOOCV yielded reasonably high rates while many of the bivariate marginals of the two calibration analyses looked alike suggests that there are indeed many-way interactions in the latent failure surface.
While we were not able to explicitly give a closed-form description of what combinations would lead to failures, our method was able to implicitly learn such conditions and incorporate them into a restricted calibration posterior.


\spacingset{1.1}     

\footnotesize

\bibliographystyle{asa}
\bibliography{0-Biblio}

\end{document}